\documentclass[aps,showpacs,showkeys,preprintnumbers]{revtex4}
\usepackage{graphicx}
\usepackage[english]{babel}  
\usepackage[usenames]{color}  
\usepackage[latin1]{inputenc}  
\usepackage{bm}  
\newcommand{\pomeron}{I\!\!P}
\begin{document}
\title{Transverse structure of strong interactions at LHC: \\
From diffraction to new particle production
\footnote{To appear in the proceedings of ``Physics at LHC'', 
Vienna, Austria, July 13--17, 2004.}}
\author{L.~Frankfurt}
\affiliation{School of Physics and Astronomy,
Tel Aviv University, Tel Aviv, 69978, Israel}
\author{M.~Strikman}
\affiliation{Department of Physics, Pennsylvania State University,
University Park, PA  16802, U.S.A.}
\author{C.~Weiss}
\affiliation{Theory Group, Jefferson Lab, Newport News, VA 23606, U.S.A.}
\author{M.~Zhalov} 
\affiliation{Petersburg Nuclear Physics Institute, Gatchina, 188300 Russia}
\begin{abstract}
We discuss the global structure of $pp$ events at LHC with 
hard processes (particle production in two--parton collisions)
on the basis of the transverse spatial characteristics of the partonic 
initial state. Studies of hard exclusive processes in $ep$ scattering
have shown that the transverse area occupied by partons with $x \geq 10^{-2}$
is much smaller than the size of the nucleon as it appears in generic 
inelastic $pp$ collisions at high energies (``two--scale picture''). 
We show that this is consistent with the observation that the elastic $pp$ 
amplitude at the Tevatron energy is close to the black body limit at small 
impact parameters. Our picture implies that inclusive heavy 
particle production (Higgs, SUSY) happens only in central $pp$ collisions.
At LHC energies, the final state characteristics of such events
are strongly influenced by the approach to the black body limit,
and thus may differ substantially from what one expects based on
the extrapolation of Tevatron results. Our two--scale picture also
allows us to analyze several types of hard diffractive processes observable 
at LHC: {\it i)}~Diffractive proton dissociation into three jets, 
which probes small--size configurations in the proton wave function; 
{\it ii)}~exclusive diffractive Higgs production, in which we estimate 
the rapidity gap survival probability; 
{\it iii)}~inclusive diffractive processes.
\end{abstract}
\pacs{12.38.-t, 13.85.-t, 14.80.Bn}
\keywords{diffraction, generalized parton distributions, central collisions}
\preprint{JLAB-THY-04-311}
\maketitle
\section{Introduction}
Hard processes, such as the production of high--$p_{\perp}$ dijets or heavy
particles, provide a chance to apply QCD to the theory of $pp/\bar p p$ 
collisions at high energies. The dominant mechanism for these processes are
collisions of two partons. Thanks to QCD factorization theorems,
the cross section for these processes can be represented as the product 
of the hard partonic cross section, calculable in perturbative QCD, 
and functions describing the distribution of partons in the initial state.
To calculate the total cross section for hard processes requires
knowledge only of the distributions of partons with respect to 
longitudinal momentum, which are known from inclusive DIS
and related experiments. A much more challenging problem is
to describe the properties of the hadronic final state in
events with hard processes, {\it e.g.}, the average hadron multiplicities 
in events with heavy particle production, or the probability 
for hard diffractive events with large rapidity gaps.
These problems require knowledge also of the transverse 
spatial distribution of partons in the initial state
(including its dependence on flavor and spin),
as well as parton--parton correlations.

%
%
\begin{figure}[t]
\includegraphics[width=5.5cm,height=3cm]{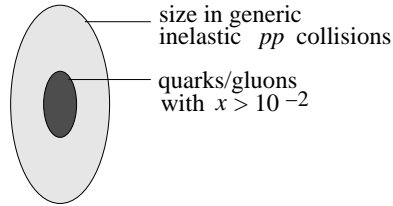}
\caption[]{The ``onion--like'' transverse structure of the nucleon
in high--energy $pp$ collisions.}
\label{fig_onion}
\end{figure}
The transverse spatial distribution of partons in the proton
is probed in hard exclusive electro/photoproduction processes,
such as the electroproduction of vector mesons ($\rho, \phi$)
or real photons (deeply virtual Compton scattering), or 
the photoproduction of heavy quarkonia ($J/\psi, \psi', \Upsilon$),
by measuring the $t$--dependence of the differential cross sections.
Such measurements at FNAL and HERA have provided a rather detailed picture
of the transverse spatial distribution of gluons down to momentum 
fractions of the order $x \sim 10^{-4}$. In particular, these
results have shown that the transverse area occupied by
partons with $x \geq 10^{-2}$ is {\it much smaller} than
the transverse size of the proton in generic inelastic 
$pp / \bar p p$ collisions at high energies.
Since the latter size rises rapidly with energy, the difference 
between the two transverse areas, which is already significant 
at Tevatron energies, is expected to become even more pronounced 
at LHC. One is dealing with an ``onion--like'' transverse structure 
of the nucleon, as depicted in Fig.~\ref{fig_onion}.
This two--scale picture is the key to understanding the properties 
of the hadronic final state in $pp$ events with hard processes at LHC. 

First, the two--scale picture of Fig.~\ref{fig_onion} implies a 
classification of $pp / \bar p p$ events in ``central'' (areas of 
large--$x$ partons overlap) and ``generic'' ones (areas of large--$x$ 
partons need not overlap). The ``generic'' collisions give the dominant 
contribution to the overall inelastic cross section. Hard processes such 
as heavy particle production will practically only happen in 
``central'' collisions \cite{Frankfurt:2003td}.
A new phenomenon expected at LHC is the approach to the ``black body limit'' 
of strong interactions due to the interactions of large--$x$ partons
with the strong small--$x$ gluon field in the other proton. This has profound 
consequences for the pattern of hadron production, in particular in the 
forward/backward rapidity region. These strong interaction effects need 
to be understood in order to reliably identify signals due 
to new particles.

Second, the two--scale picture of the transverse structure of the
nucleon can be applied to diffractive processes in $pp$ scattering, 
in which the particles produced in the hard process are separated 
from the rest by rapidity gaps. One example is the diffractive dissociation 
of one of the protons into three jets,
$p + p \rightarrow \mbox{3 jets} + \mbox{(gap)} + p$.
This process proceeds from small--size configurations in the
dissociating proton, and is the analogue of the diffractive
two--jet dissociation of pions in $\pi A$ collisions.
In particular, measurements of such processes will allow to investigate 
whether the fast increase with energy of the gluon density in the 
nucleon, predicted by DGLAP evolution in QCD and observed at HERA,
will slow down as suggested by various resummation approaches
\cite{Ciafaloni:2003rd}. Another example is the exclusive diffractive 
production of heavy particles, 
$p + p \rightarrow p + \mbox{(gap)} + H + \mbox{(gap)} + p$,
which is being discussed as a candidate for the Higgs search at 
LHC \cite{Kaidalov:2003fw}. Here, the requirement that soft interactions
between the spectator systems preserve the rapidity gaps
results in a suppression of the cross section compared to the
non-diffractive case, which can be estimated using information
from $pp$ elastic scattering. The same applies to inclusive diffractive
processes, $p + p \rightarrow p + \mbox{(gap)} + \mbox{2 jets} + X$
or $p + \mbox{(gap)} + \mbox{2 jets} + X + \mbox{(gap)} + p$. 
\section{Transverse structure of the nucleon in high--energy processes}
\label{sec_transverse}
\begin{figure}[t]
\includegraphics[width=6.7cm,height=7cm]{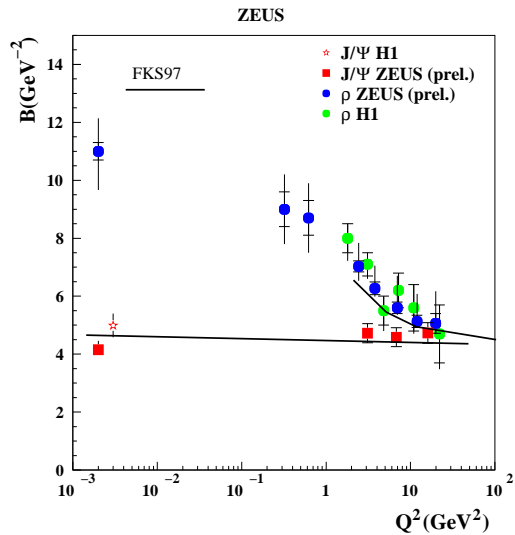}
\caption[]{Convergence of the $t$--slopes of $\rho$ and $J/\psi$ 
electroproduction at large $Q^2$. The compilation of the data is from 
Ref.~\cite{Levy:2003qz}. The solid curves are the theoretical predictions of 
Ref.~\cite{Frankfurt:1997fj} for the $Q^2$ dependence of the $t$--slopes.}
\label{fig_levy}
\end{figure}

Information about the spatial distribution of partons in the transverse 
plane comes from studies of hard exclusive processes in $ep$ scattering, 
in particular the electroproduction of light vector mesons 
($\rho,\phi$) at large $Q^2$, and the photoproduction of heavy quarkonia 
($J/\psi, \psi^{\prime},\Upsilon$). These processes probe
the generalized parton distributions in the nucleon. 
The extensive data on vector meson production at HERA (for a review, 
see Ref.\cite{Abramowicz:1998ii}) support the dominance of the partonic 
mechanism. In particular, they confirm a qualitative prediction of QCD 
factorization --- the convergence of the $t$--slope of $\rho$ meson 
electroproduction at large $Q^2$ to the slope of $J/\psi$ 
photo/electroproduction \cite{Brodsky:1994kf}, see Fig.~\ref{fig_levy}. 
Numerical calculations, accounting for the finite transverse size of 
the wave function of the produced vector meson, explain the key features 
of the high--$Q^2$ data \cite{Frankfurt:1997fj}. The $t$--dependence of 
the cross section for $J/\psi$ photo/electroproduction is given 
(up to small corrections) by the square of the ``two--gluon form factor'' 
of the nucleon \cite{Frankfurt:2002ka},
\begin{equation}
\frac{d\sigma^{\gamma^* p \rightarrow J/\psi p}}{dt}(t) 
\;\; \propto \;\; \left[ F_g (x, t) \right]^2 ,
\hspace{4em} x \, \sim \, M_{\bar c c}^2 / W^2 .
\end{equation}
The two--dimensional Fourier transform of this function,
\begin{equation}
F_g (x, \rho ) \;\; \equiv \;\; \int \frac{d^2 \Delta_\perp}{(2\pi )^2} \; 
e^{-i (\bm{\Delta}_\perp \bm{\rho})} \; F_g (x, t)
\hspace{4em}
(t = -\bm{\Delta}_\perp^2 ), 
\label{impact_def}
\end{equation}
describes the spatial distribution of gluons with momentum 
fraction $x$ in the transverse plane [the distribution is
normalized as $\int d^2 \rho \, F_g (x, \rho) = 1$]. Thus, one can
map the distribution of gluons in the transverse plane from
the measured $t$--dependence of the cross section.
%
%

A simple phenomenological parameterization of the two--gluon form
factor of the nucleon, including its variation with $x$,
is the dipole form \cite{Frankfurt:2002ka,Frankfurt:2003td}
\begin{equation}
F_g (x, t) \;\; = \;\; (1 - t/m_g^2)^{-2},
\hspace{4em} m_g^2 \;\; \equiv \;\; m_g^2 (x, Q^2) .
\label{dipole}
\end{equation}
This form is motivated by the expectation that at $x \geq 0.1$,
where the contribution of the nucleon's pion cloud is suppressed,
the two--gluon form factor should follow the axial form factor
of the nucleon ($m_g^2 \approx m_{A}^2 = 1.1\, {\rm GeV}^2$).
It describes well the $t$--slopes of $J/\psi$ photoproduction
measured in a number of fixed--target experiments \cite{Frankfurt:2002ka}.
The experimentally observed increase of the slope between 
$x \sim 0.1$ and $x \sim 0.01$ can be explained by contributions
of the nucleon's pion cloud to the gluon distribution at transverse
distances $\rho \sim 1/(2 M_\pi)$ \cite{Strikman:2003gz}.
This effect, as well as the subsequent slow increase of the slope at 
smaller $x$, and the small effects of DGLAP evolution, are incorporated
by way of an $x$-- and $Q^2$--dependence of the dipole mass parameter,
see Refs.~\cite{Frankfurt:2003td,Strikman:2004km} for details. The Fourier 
transform of Eq.~(\ref{dipole}), {\it cf.}\ Eq.~(\ref{impact_def}), defines 
our parametrization of the transverse spatial distribution of gluons. 

A crucial observation is that the transverse area occupied by partons
with significant momentum fraction, $x \geq 10^{-2}$, is much
smaller than the size of the nucleon as it appears in generic
inelastic $pp$ collisions at high energies. This can qualitatively
be understood in Gribov's picture of the partonic wave function
of the nucleon, which allows for a unified description of 
electromagnetic and strong interactions at high energies \cite{Gribov:1973jg}.
In this picture, the $pp$ cross section is dominated by collisions
of ``soft'' partons (very small $x$), which are the result of
successive decays of the ``hard'' partons at large $x$.
This decay process has the character of a random walk in the
transverse plane (``Gribov diffusion''). Each decay process
results in a shift of the transverse position inversely proportional 
to the transverse momentum in the decay vertex, $\sim 1/k_{\perp}$.
Due to the randomness of the emissions, and the constant degrading 
of the longitudinal momentum, one finds that $n$ successive decays
result in an overall increase of the transverse area of
\begin{equation}
R^2(n) \;\; \propto \;\; \frac{n}{k_{\perp}^2} \;\; \propto \;\; 
\frac{\ln (x_0/x)}{k_{\perp}^2}. 
\end{equation}
The transverse area occupied by soft partons thus grows linearly with $\ln s$.
For hard partons the growth of the transverse area with the collision
energy is much weaker, due to the smaller scale of the random shifts 
in impact parameter space. In addition, the transverse area occupied 
by the hard partons at large $x$ is already small. This leads to the
appearance of two widely separated transverse area scales in $pp$
collisions at high energies (see Fig.~\ref{fig_onion}) --- a small
transverse area occupied by hard partons, and a much larger
area occupied by soft partons.

\begin{figure}[t]
\begin{tabular}{cc}
\includegraphics[width=7cm,height=7cm]{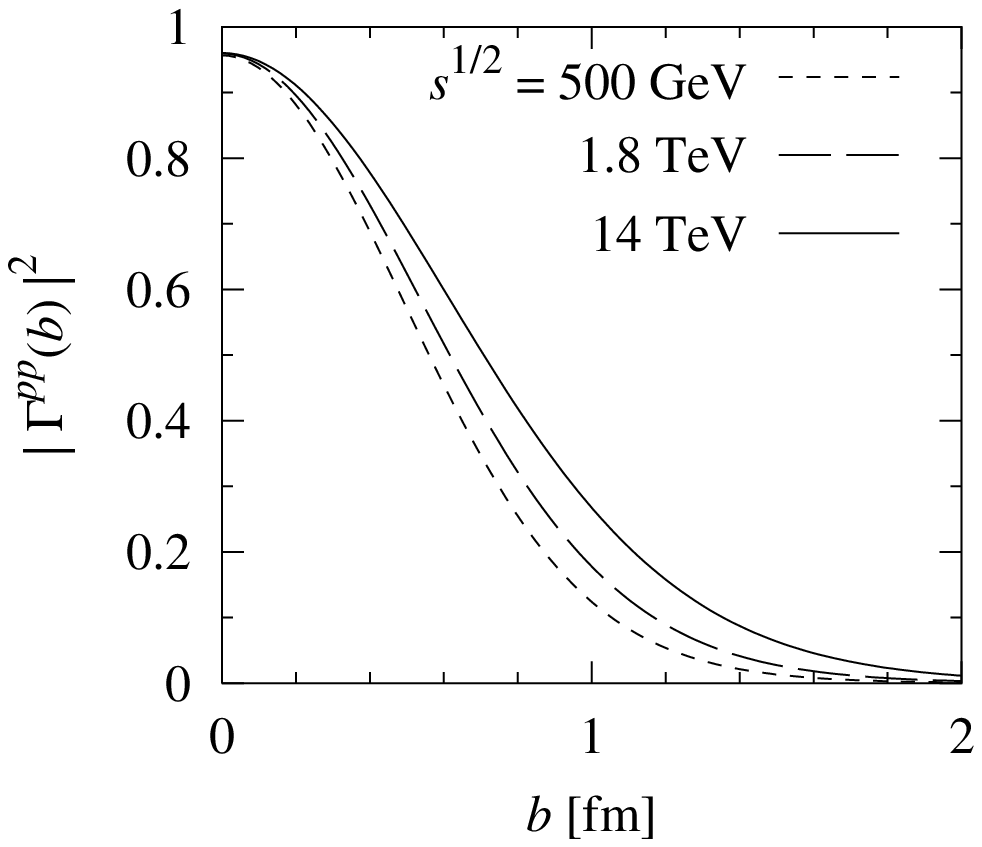}
&
\includegraphics[width=7cm,height=7cm]{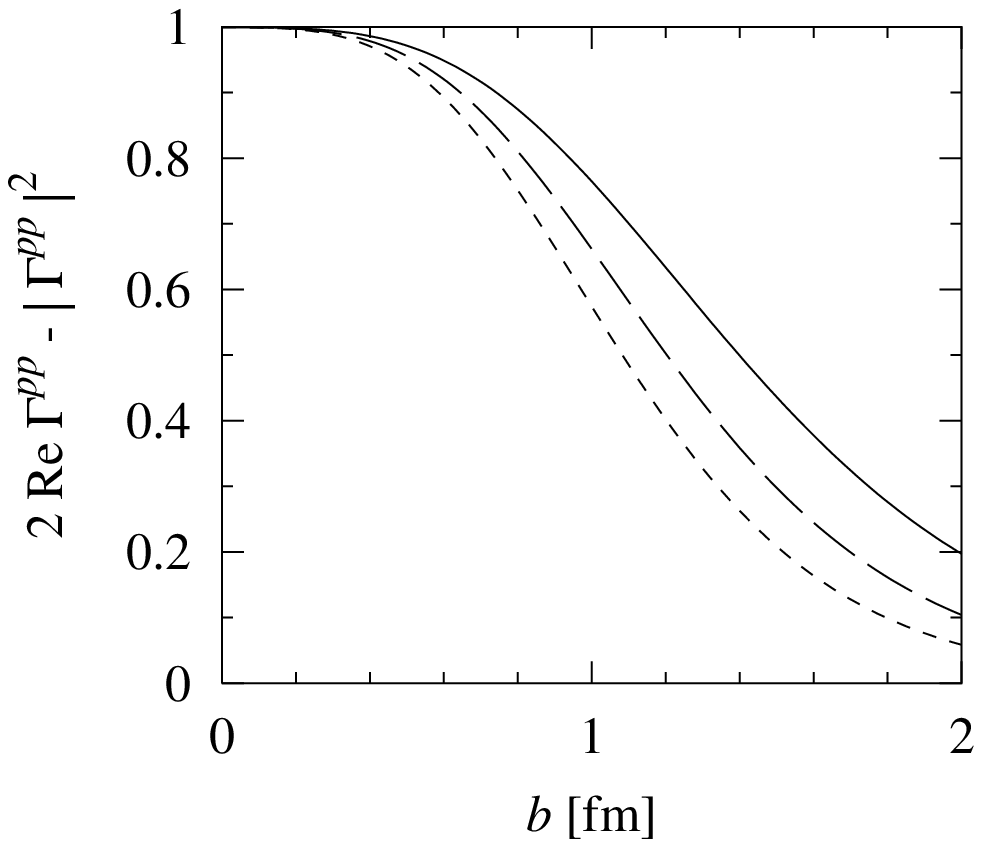}
\end{tabular}
\caption[]{Left panel: Modulus squared of the $pp$ elastic amplitude
in impact parameter representation, $\Gamma^{pp}(b)$, Eq.~(\ref{Gamma}).
Right panel: The probability for inelastic interaction at 
a given impact parameter, Eq.~(\ref{winel}).}
\label{fig_Gamma}
\end{figure}
In order to quantify the transverse size of the nucleon 
in inelastic $pp$ collisions one can use phenomenological 
parameterizations of the $pp$ elastic scattering
amplitude at high energies in the impact parameter representation, 
\begin{equation}
\Gamma^{pp}(s,b) \;\; \equiv \;\; \frac{4\pi}{is} \int 
\frac{d^2 \Delta_\perp}{(2\pi)^2} \; e^{-i (\bm{\Delta}_\perp \bm{b})}
A^{pp} (s,t)
\hspace{4em}
(t = -\bm{\Delta}_\perp^2 ).
\label{Gamma}
\end{equation}
By unitarity, the probability for an inelastic collision at a 
given impact parameter, $b$, is related to $\Gamma^{pp} (s, b)$ as 
\begin{equation}
2\, \mbox{Re} \, \Gamma^{pp}(s,b) - |\Gamma^{pp}(s,b)|^2.
\label{winel}
\end{equation}
Fig.~\ref{fig_Gamma} shows $|\Gamma^{pp}(s,b)|^2$ and 
the probability (\ref{winel}) obtained
from the parameterization of Ref.~\cite{Islam:2002au}. One clearly sees 
how the radius of strong interactions increases with the collision energy. 
In Gribov's picture, this reflects the increase of the transverse 
area occupied by the soft partons. 
\section{Hard interactions and the ``black body limit'' in elastic $pp$ 
scattering}
\label{sec_hardblack}
Parameterizations of the $pp$ elastic amplitude indicate 
that at high energies the strength of the interaction 
at small impact parameters approaches the ``black body limit'', 
$\Gamma^{pp} \approx 1$, corresponding to unit probability for inelastic 
interactions, see Fig.~\ref{fig_Gamma}. A rough estimate of the maximum 
impact parameter for which the interaction remains black can be made 
in the single Pomeron exchange model of the total cross section.
The minimum invariant energy for which $\Gamma^{pp} (b = 0) \approx 1$
is approximately the Tevatron energy, $s_T \approx (2\, \mbox{TeV})^2$.
For $s > s_T$, the total cross section then grows as
$\sigma_{\rm tot} \propto (s/s_T)^{\alpha_{\pomeron}(0)-1}$, 
while the $t$--slope of the elastic cross section grows as
$B_{\pomeron} = B_T + 2  \alpha'\ln(s/s_T)$. One finds that 
$\Gamma^{pp}$ reaches unity at
\begin{equation}
b_F^2 \;\; \approx \;\; \left[ \alpha_{\pomeron}(0)-1 \right]
\ln (s/s_T) \; B_{\pomeron} .
\label{cutoff}
\end{equation}
This corresponds to the universal Froissart limiting behavior of
hadronic total cross sections (same for all hadrons and nuclei).

In QCD, the ``blackening'' of the $pp$ interaction at small impact
parameters could arise from hard as well as soft interactions.
It is interesting to estimate the role of hard interactions 
in the approach to the black body limit within our two--scale 
picture of the transverse partonic structure of the nucleon.
The analysis of Ref.~\cite{Frankfurt:2003td} 
(see Section~\ref{sec_central} below) shows that in central 
$pp$ collisions the leading quarks and gluons receive significant transverse 
momenta when passing through the gluon field of the other proton. 
For example, at Tevatron energies, quarks with $x_1 \ge 0.2$ 
get on average transverse momenta $> 1 \, {\rm GeV}/c$,
see Section~\ref{sec_bbl} below. When a single leading quark gets 
a transverse momentum, $p_{\perp}$, the probability for the 
nucleon to remain intact is approximately given by the square of the 
nucleon form factor, $F_N^2(p_{\perp}^2)$, which is $\le 0.1 $ for 
$p_{\perp}> 1 \, {\rm GeV}/c$, {\it i.e.}, very small. One may thus conclude
that the probability of the survival averaged over $p_{\perp}$ 
should be less than 1/2 (on average, half of the time the quark should
receive a transverse momentum larger than the average transverse momentum, 
$p_{\perp}> 1 \, {\rm GeV}/c$). Since there are six leading quarks (plus a 
number of leading gluons), the probability for the two protons 
to stay intact, $|1 - \Gamma^{pp}(s, b)|^2$, should go as a high power of the 
survival probability for the case of single parton removal, 
and thus be very small. 
This crude estimate shows that already at Tevatron energies
$|1 - \Gamma^{pp}(s, b = 0)|^2$ should be close to zero due to hard 
interactions, which is consistent with phenomenological parameterizations
of the $pp$ elastic amplitude, see above. We conclude that hard interactions 
alone are sufficient to ensure ``blackness'' of the inelastic interactions 
at small impact parameters.

\begin{figure}[t]
\includegraphics[width=8.6cm,height=6cm]{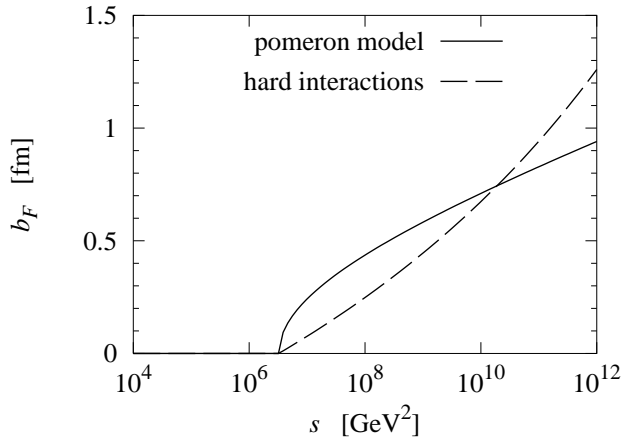}
\caption[]{Energy dependence of the maximum impact parameter for
``black'' $pp$ interactions, $b_F$. Solid line: Estimate based on 
the pomeron model, Eq.~(\ref{cutoff}). Dashed line: Estimate based
on hard interactions, Eq.~\ref{pwave4}.}
\label{fig_bf}
\end{figure}
One can estimate the maximum impact parameter up to which
hard interactions cause the $pp$ interaction to be ``black''. 
The probability for a leading parton with $x_1 \sim 10^{-1}$
to experience a hard inelastic interaction increases with the 
collision energy at least as fast as corresponding to the increase 
of the gluon density in the other proton at
$x_2 = 4 p_\perp^2 / (s x_1)$, {\it cf.}\ Eq.~(\ref{x_resolved}) below.
Since $x_2 G_N(x_2,Q^2) \propto x_2^{-n_h}$ 
with $n_h\ge 0.2$, this implies that the probability should grow as 
$s^{n_h}$ \footnote{The HERA data on dipole--nucleon scattering suggest 
that the taming of the gluon density starts only when the probability of 
inelastic interactions becomes large, $\ge 1/2$. However, for such 
probabilities of single parton interactions, multiparton interactions 
insure that the overall interaction is practically black.}. 
Our parametrization of the transverse spatial distribution of gluons, 
{\it cf.}\ Eq.~\ref{impact_def}, suggests
that the gluon density decreases with the distance from 
the center of the nucleon approximately as 
$\sim \exp \left[ -m_g(x_2 )\rho \right]$. If one neglects the transverse 
spread of the large--$x_1$ partons as compared to that of the small--$x_2$ 
gluons one arrives at an estimate of the energy dependence of 
$b_F$ as due to hard interactions \cite{Frankfurt:2004fm},
\begin{equation}
b_F \;\; \approx \;\; \frac{n_h \ln (s/s_T)}{m_g (x_2 )} ,
\label{pwave4}
\end{equation}
with $x_2$ as defined above. It is very encouraging that this estimate 
agrees well with that obtained above in the pomeron model, 
Eq.~(\ref{cutoff}), see Fig.~\ref{fig_bf}. This shows that our two--scale 
picture of the transverse structure of the nucleon --- and the ensuing
picture of hard and soft interactions --- are self--consistent.

Measurements of elastic $pp$ scattering at LHC will allow to 
investigate the onset of the black body limit in $pp$ interactions.
Such studies not only probe the structure of the nucleon periphery 
at very high energies, but also provide important constraints 
for models of inelastic interactions at smaller impact parameters.
\section{New particle production in central $pp$ collisions}
\label{sec_central}
\begin{figure}[t]
\includegraphics[width=8cm,height=5.4cm]{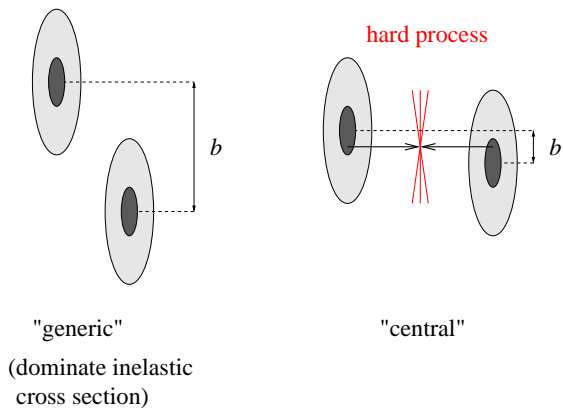}
\caption[]{Classification of $pp$ events at high energies.}
\label{fig_coll}
\end{figure}

The two--scale picture of the transverse structure of the 
nucleon, see Fig.~\ref{fig_onion} and Section~\ref{sec_transverse},
implies a classification of $pp / \bar p p$ events in ``central'' 
(areas of large--$x$ partons overlap) and ``generic'' ones 
(areas of large--$x$ partons need not overlap),
see Fig.~\ref{fig_coll} \cite{Frankfurt:2003td}. 
Generic collisions give the dominant contribution to 
the overall inelastic cross section.
Hard processes, such as heavy particle production at central
rapidities, will practically only happen in central collisions.

\begin{figure}[b]
\begin{tabular}{cc}
\includegraphics[width=7cm,height=7cm]{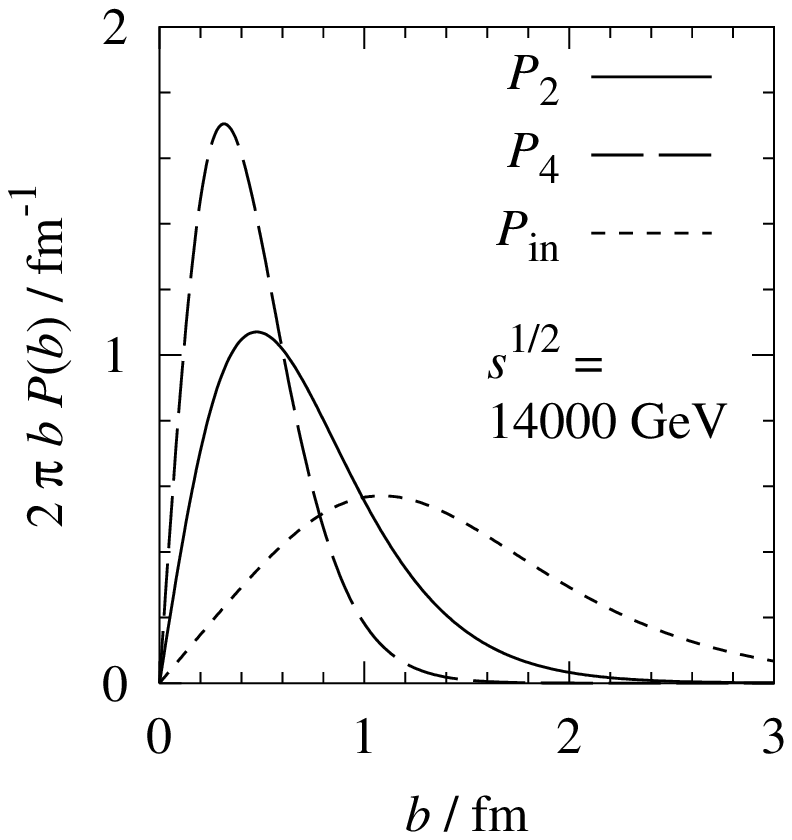}
&
\includegraphics[width=7cm,height=7cm]{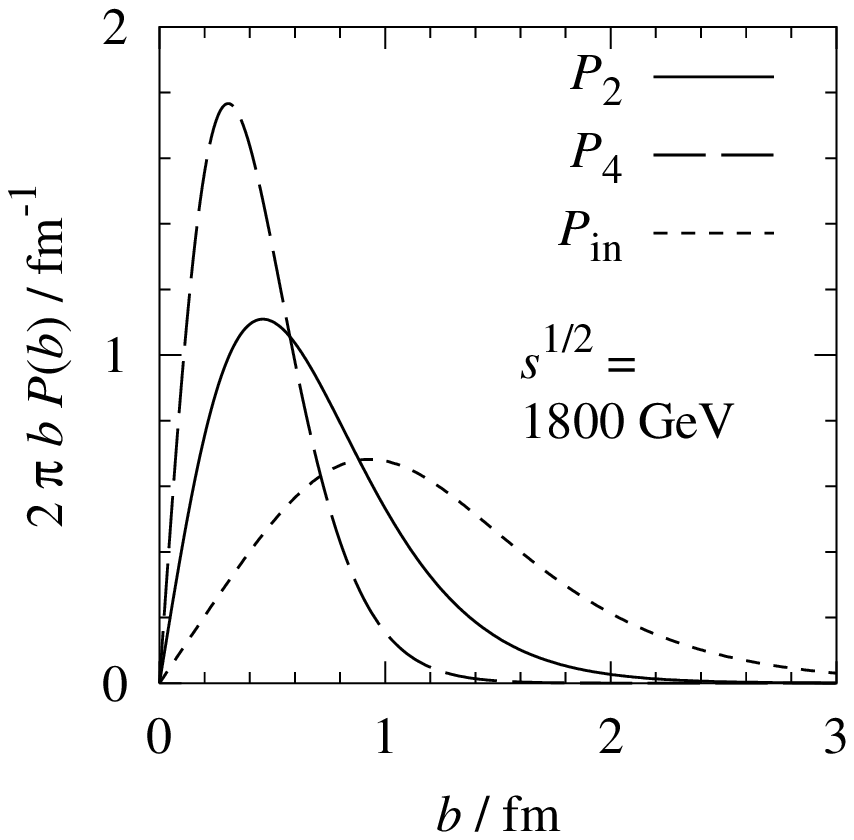}
\end{tabular}
\caption[]{\textit{Solid lines:} $b$--distributions for events
with hard dijet production, $P_2 (b)$, with $q_\perp = 25 \, \text{GeV}$.
\textit{Long--dashed line:} $b$--distribution for double dijet events, 
$P_4 (b)$. \textit{Short--dashed line}: $b$--distribution 
for generic inelastic collisions, obtained from the parameterization 
of the elastic $pp$ amplitude of Islam \textit{et al.} \cite{Islam:2002au}.
The plots show the ``radial'' distributions in the impact parameter
plane, $2 \pi b \, P (b)$.} 
\label{fig_pb}
\end{figure}

To quantify the distinction between generic and central collisions, 
we estimate the distribution of both types of events over 
the $pp$ impact parameters. For generic collisions, the distribution 
is determined by the $b$--dependent probability of inelastic interaction, 
Eq.~(\ref{winel}), obtained via unitarity from the elastic $pp$ amplitude
in the impact parameter representation. We define a normalized
$b$--distribution as
\begin{equation}
P_{\text{in}} (s, b) \;\; = \;\; 
\frac{2 \text{Re}\; \Gamma^{pp} (s, b) - |\Gamma^{pp} (s, b)|^2}
{\sigma_{\text{in}} (s)} ,
\label{P_in_def}
\end{equation}
where $\sigma_{\text{in}}(s)$ is the inelastic cross section, which is
given by the integral $\int d^2 b$ of the expression in the numerator.
For collisions with a hard process, on the other hand,
the $b$--distribution is determined by the overlap integral of the
distribution of hard partons in the two colliding protons
(see Fig.~\ref{fig_coll}),
\begin{equation}
P_2 (b) \;\; \equiv \;\; \int d^2\rho_1 \int d^2\rho_2 \; 
\delta^{(2)} (\bm{b} - \bm{\rho}_1 + \bm{\rho}_2 )
\; F_g (x_1, \rho_1 ) \; F_g (x_2, \rho_2 ) .
\label{P_2}
\end{equation}
Numerical estimates can be performed with our parametrization of the
transverse spatial distribution of hard gluons, Eq.~(\ref{dipole}),
which takes into account the change of the distribution with $x$
and the scale of the hard process. The two $b$--distributions are compared
in Fig.~\ref{fig_pb} for Tevatron and LHC energies. For the hard process 
we have taken the production of a dijet with $q_\perp = 25\,{\rm GeV}$;
in the case of Higgs production at LHC the distribution $P_2 (b)$
would be even narrower. The results clearly show that events with
hard processes have a much narrower impact parameter distribution
than generic inelastic events.
%
%
%
%
%

One expects that at LHC energies the rate of production 
of two pairs of jets will be very high. It is interesting to consider 
the $b$--distribution also for the production of two dijets in two 
binary parton--parton collisions. Neglecting possible correlations 
between the partons in the transverse plane it is given by
\begin{equation}
P_4 (b) \;\; \equiv \;\; \frac{\left[ P_2 (b)\right]^2}
{\int d^2 b \; \left[ P_2 (b)\right]^2} .
\label{P_4}
\end{equation}
Fig.~\ref{fig_pb} shows that this distribution is significantly
narrower than $P_2$, {\it i.e.}, the requirement of two hard processes
results in a further reduction of effective impact parameters.
Note that there are indications of significant correlations in the 
transverse positions of two partons, which would effectively reduce the 
difference between the $b$--distribution in double dijet 
and single dijet events; see Refs.~\cite{Frankfurt:2003td,Frankfurt:2004kn} 
for details. 
\section{Onset of the black body limit in central $pp$ collisions}
\label{sec_bbl}
A new phenomenon encountered in central $pp$ collisions
at LHC energies is that the interaction of leading partons
in one proton with the gluon field in the other proton approaches 
the maximum strength allowed by $s$-channel unitarity
(``black body limit''). As a result, the structure of the
final state in central $pp$ collisions differs in many respects
--- leading particle spectra, multiplicities of soft hadrons 
at central rapidities, long range correlations --- from that
of generic inelastic events.

\begin{figure}[b]
\includegraphics[width=6.4cm,height=2.5cm]{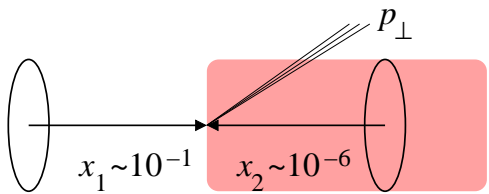}
\caption[]{The black body limit in central $pp$ collisions:
Leading partons in one proton, $x_1 \sim 10^{-1}$, interact
with a dense medium of small--$x_2$ gluons in the other proton (shaded area),
acquiring a large transverse momentum, $p_{\perp}$.}
\label{fig_bbl}
\end{figure}
Generally, in particle production in hard parton--parton collisions,
a parton with momentum fraction $x_1$ and transverse momentum
$p_{\perp}$ in one proton resolves partons in the other proton 
with momentum fraction
\begin{equation}
x_2 \;\; = \;\; \frac{4 \, p_\perp^2}{x_1 \, s} .
\label{x_resolved}
\end{equation}
In leading particle production at LHC ($\sqrt{s} = 14 \,\text{TeV}$) 
a leading parton with $x_1 \sim 10^{-1}$ and significant transverse
momentum, $p_\perp \sim 2\, \text{GeV}/c$, resolves partons with 
$x_2 \sim 10^{-6}$ in the other proton. In this situation the gluon 
density in this proton is so large that ``taming'' effects cannot 
be neglected; for a review and references see Ref.~\cite{Frankfurt:2000ty}. 
The leading parton can be thought of as propagating through a high--density 
``gluon medium'', or, equivalently, a strong gluon field, in the other proton,
see Fig.~\ref{fig_bbl}. The interaction of the leading parton with the 
other proton approaches the ``black body limit'', characterized 
by a near unity probability for inelastic scattering. In fact, 
in $pp$ collisions at LHC the leading partons propagate through gluon 
fields with a strength comparable to that in central $pA$ collisions 
at similar energies.

\begin{figure}[t]
\begin{tabular}{cc}
\includegraphics[width=7cm,height=7cm]{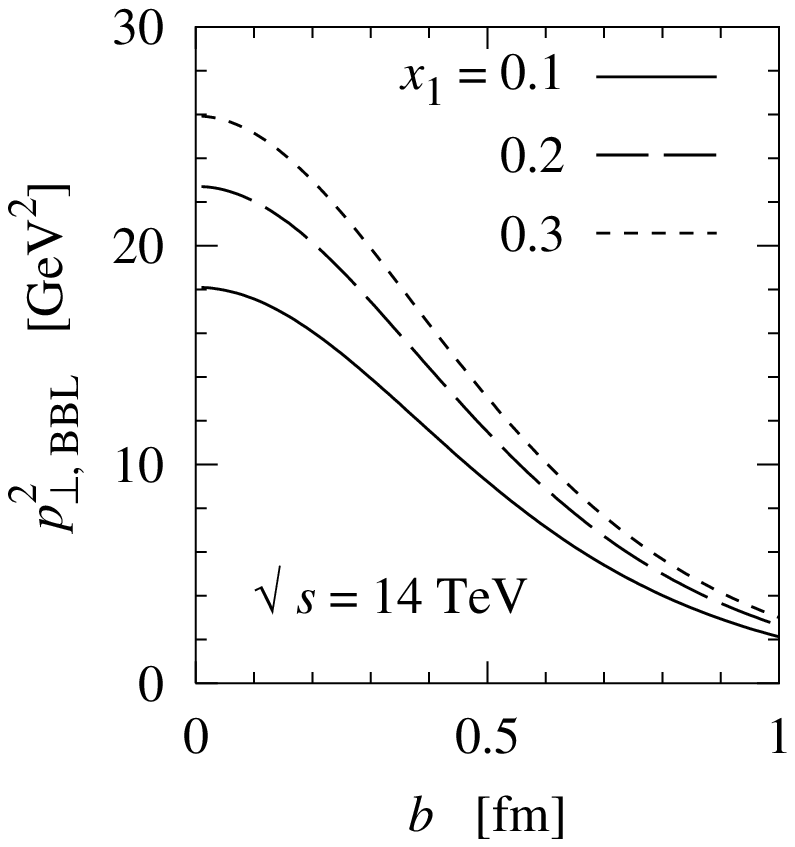}
&
\includegraphics[width=7cm,height=7cm]{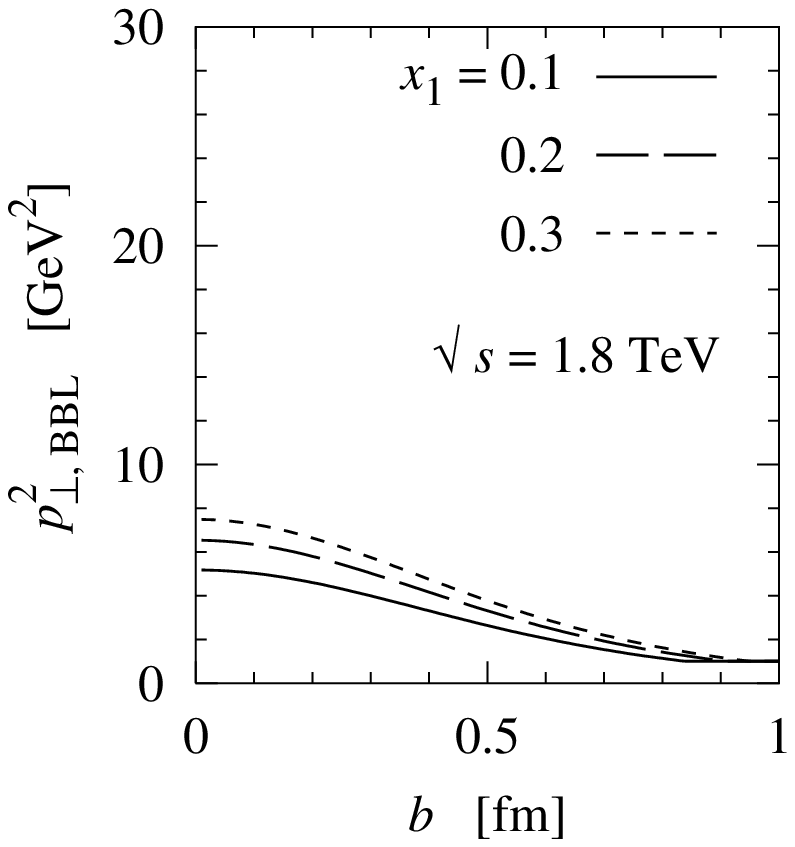}
\end{tabular}
\caption[]{The critical transverse momentum squared, 
$p_{\perp , \text{BBL}}^2$, below which the interaction of a
leading gluon (momentum fractin $x_1$) with the other
proton is close to the black body limit, as a function of 
the impact parameter of the $pp$ collision, $b$. For leading 
quarks, the values of $p_{\perp , \text{BBL}}^2$ are about
half of those for gluons shown here. Shown are the estimates
for LHC (left panel) and Tevatron energies (right panel).}
\label{fig_pt}
\end{figure}

The ``black'' interactions are experienced by leading partons
with transverse momenta up to a certain critical value, 
$p_{\perp,\text{BBL}}$ [the $x_2$ of the gluon density probed 
is proportional to $p_{\perp}^2$, {\it cf.}\ Eq.~(\ref{x_resolved})].
Conversely, this means that low--$p_{\perp}$ leading partons
will on average acquire transverse momenta of the order
$p_{\perp,\text{BBL}}$ through their interactions with the
strong gluon field, which results in certain modifications
of the pattern or particle production at forward/backward rapidities, 
see Section~\ref{sec_final} below.\footnote{The kinematics of the final
state produced in the interaction of the large--$x_1$ parton with 
the small--$x_2$ gluon field resembles the backscattering of a laser beam 
off a high--energy electron beam. The large--$x_1$ parton gets a significant 
transverse momentum and loses a certain fraction of its longitudinal 
momentum, accelerating at the same time a small--$x_2$ parton.} 
We have estimated the critical transverse momentum, $p_{\perp,\text{BBL}}$, 
using phenomenological parametrizations of the gluon density at small $x_2$,
and our parametrization of the transverse spatial distribution of 
gluons, Eq.~(\ref{dipole}) \cite{Frankfurt:2003td}. 
For simplicity, we considered the scattering of a small color--singlet 
dipole off the ``target'' nucleon, in the spirit of the dipole picture of
high--energy scattering in the BFKL model \cite{Mueller:1994jq},
and then converted the kinematical variables to those corresponding
to a leading parton in proton--proton scattering. 
Our criterion for proximity to the black body limit (BBL) was that 
$\Gamma^{\text{dipole-proton}} (s, b) > 0.5$,
which corresponds to a probability for inelastic interaction of $>0.75$.
Fig.~\ref{fig_pt} shows an estimate of the critical transverse momentum 
for leading partons with momentum fractions of the order $x_1 \sim 10^{-1}$, 
as a function of the impact parameter of the $pp$ 
system \cite{Frankfurt:2003td}. Because the gluon density at 
$x_2 \sim 10^{-6}$ is maximum in the transverse center of the proton, 
and the distribution of leading partons is likewise concentrated 
at small transverse distances, the chances for ``black'' interactions
are greatest in central $pp$ collisions (as {\it e.g.} those in 
which heavy particles are produced). This is reflected in the rapid
drop of $p_{\perp,\text{BBL}}^2$ with $b$ in Fig.~\ref{fig_pt}.
One sees that $p_{\perp,\text{BBL}} \sim \mbox{several} \; {\rm GeV}$
in central collisions at LHC. Substantially smaller values are 
obtained at the Tevatron energy.

\begin{figure}[t]
\begin{tabular}{cc}
\includegraphics[width=7cm,height=7cm]{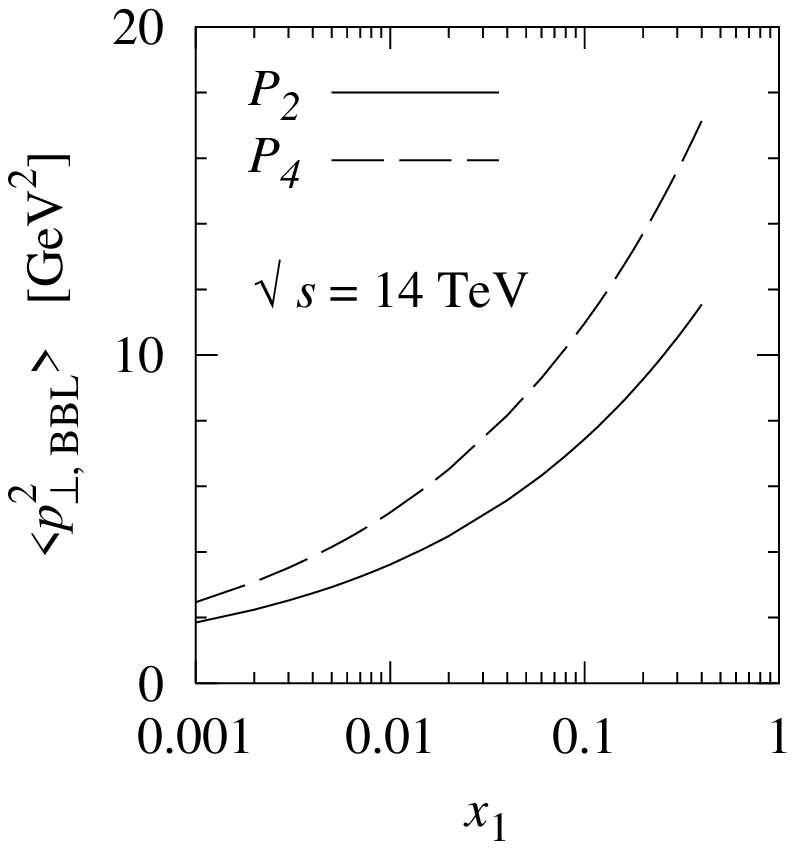}
&
\includegraphics[width=7cm,height=7cm]{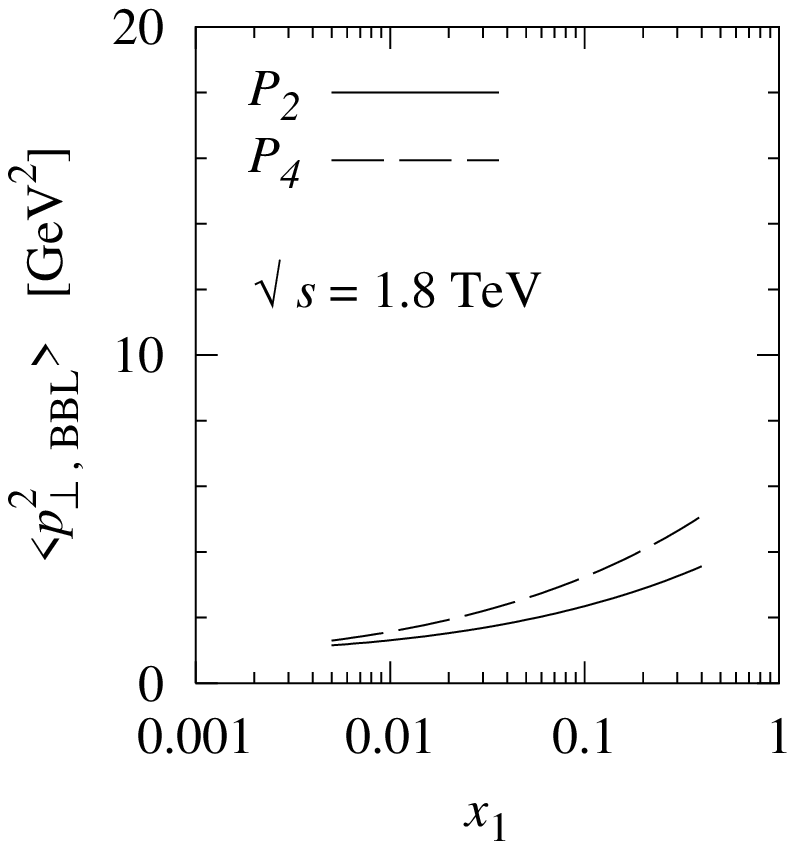}
\end{tabular}
\caption[]{The average value of $p_{\perp , \text{BBL}}^2$, 
\textit{cf.}\ Fig.~\ref{fig_pt}, over impact parameters of the 
$pp$ collision, as a function of the leading gluon's momentum 
fraction, $x_1$. The averages were computed with the impact 
parameter distribution corresponding to the hard dijet trigger, 
$P_2$ (solid line), and the double dijet trigger, $P_4$ (dashed line), 
{\it cf.}\ Fig.~\ref{fig_pb}. For leading quarks, the values of 
$\langle p_{\perp , \text{BBL}}^2 \rangle$ are about half of those
for gluons shown here. Shown are the estimates
for LHC (left panel) and Tevatron energies (right panel).}
\label{fig_ptav}
\end{figure}

To determine the typical transverse momenta of leading partons
in events with new particle (or hard dijet) production,
we need to average the results for $p_{\perp,\text{BBL}}^2$,
Fig.~\ref{fig_pt}, over $pp$ impact parameters, with the 
distribution implied by the hard production process, $P_2(b)$, 
Eq.~(\ref{P_2}) [or, in the case of four jet production, with $P_4(b)$, 
Eq.~(\ref{P_4})]. We find that the suppression of large impact parameters
implied by the hard process ({\it cf.}\ Fig.~\ref{fig_coll}) is 
sufficient to keep $p_{\perp,\text{BBL}}$ above 
$1\, {\rm GeV}/c$ in more than 99\% of events at LHC. The resulting 
average values of $p_{\perp,\text{BBL}}^2$ are shown in 
Fig.~\ref{fig_ptav}.

Our estimates show that in $pp$ collisions at LHC the leading partons 
acquire substantial transverse momenta due to ``black interactions'' 
with the small--$x_2$ gluons in the other proton. A much weaker effect 
is found at the Tevatron energy. The origin of this difference is
the increase in the gluon density due to the decrease of $x_2$ between 
Tevatron and LHC energies, {\it cf.}\ Eq.~(\ref{x_resolved}).
This indicates that the final state characteristics of central $pp$
collisions at LHC will be substantially different from what one
expects from the extrapolation of Tevatron results, see 
Section~\ref{sec_final}. 

Finally, the large values of the critical transverse momentum for ``black
interactions'' of leading partons at LHC confirm our assessment 
of Section~\ref{sec_hardblack}, that the probability for two nucleons 
at LHC energies not to interact inelastically at small impact parameters 
is very small.
\newpage
\section{Final state properties in central $pp$ collisions}
\label{sec_final}
In central $pp$ collisions the leading partons acquire large transverse
momenta, $\sim p_{\perp,\text{BBL}}$, due to the approach to the
black body limit. As a result, particle production at forward/backward 
rapidities will largely be dominated by incoherent parton fragmentation. 
One thus expects the following modifications compared to generic 
inelastic collisions:
\begin{itemize} 
\item 
Strong suppression of leading particles, in particular nucleons 
(for $z\ge 0.1$ the differential multiplicity of pions should exceed 
that of nucleons)
\item Average transverse momenta of the leading particles 
$\geq 1\, \text{GeV}/c$
\item No correlations between the transverse momenta of leading hadrons
(some correlations will remain, however, because two partons produced
in collisions of large--$x_1$ and small--$x_2$ partons may end up 
at similar rapidities)
\item 
A large fraction of events with no particles with $z\ge 0.02
\div 0.05$ in both fragmentation regions (emergence of long--range rapidity 
correlations), and large energy release at rapidities $y = 4 \div 6$.
\end{itemize}

Another effect of the multiple scattering of large--$x_1$ partons
with small--$x_2$ gluons in the other proton is the shift
of a large number of small--$x_2$ gluons to larger rapidities
(``holes''). This amounts to the creation of a substantial amount 
of color charge, and should result in an increase of soft particle 
multiplicities over a broad range of rapidities 
(albeit much smaller than if the particles originated from 
independent fragmentation). Such an increase should in fact be
present already at Tevatron energies, in events with a trigger on
two--jet or $Z^0$ production. An increase of the multiplicity at 
rapidities $|y|\le 1.0$ was indeed observed in 
Ref.~\cite{Field:2002vt}, which investigated the correlation of the
underlying event structure with the presence of such a trigger. 
It is important to extend these studies to higher rapidities.

To summarize, in $pp$ collisions at LHC 
new particles will be produced in a much more ``violent'' 
strong--interaction environment than one would expect from 
the extrapolation of the properties of minimum bias events 
at the Tevatron. Even the extrapolation of properties of 
hard dijet events should not be smooth, as the transverse 
momenta acquired by leading partons are estimated to be
substantially larger at LHC than at Tevatron, see
Figs.~\ref{fig_pt} and \ref{fig_ptav}.
\section{Diffractive proton dissociation into three jets}
\label{3jetsec}

LHC will offer an opportunity to study a variety of hard 
diffractive processes in $pp$ and $pA$ scattering. One interesting 
aspect of such processes is that they allow to probe rare small--size 
configurations in the nucleon wave function.

Recently, a color transparency phenomenon was observed in the coherent
dissociation of pions into two high--$p_\perp$ jets,
$\pi + p (A) \rightarrow \mbox{jet1} + \mbox{jet2} + p(A)$  
\cite{Aitala:2000hc}, consistent with the predictions of 
Refs.~\cite{Frankfurt:it,Frankfurt:2000jm}. 
In this process the pion scatters off the target in a point--like 
$q\bar q$ configuration. Similarly, in the proton wave function there
should exist configurations consisting of only valence quarks, and having a 
small transverse size. A proton in such a configuration can scatter 
elastically off the target and fragment into three jets, corresponding to the 
process
\begin{equation}
p + p (A) \;\; \rightarrow \;\; 
\mbox{jet1} + \mbox{jet2} + \mbox{jet3} + p(A).
\label{pp3jets}
\end{equation}

The cross section for the diffractive process (\ref{pp3jets}) 
can be evaluated based on the kind of QCD factorization theorem derived in 
Ref.~\cite{Frankfurt:2000jm}. It is proportional to the square of the 
gluon density in the nucleon at $x\approx M^2 (\mbox{3 jets})/ s$, 
and virtuality $Q^2 \sim (1\div 2) \, p_\perp^2$ \cite{Frankfurt:1998eu}.
The distribution over the fractions of the proton longitudinal momentum 
carried by the jets is proportional to the square of the light--cone 
wave function of the $|qqq \rangle$ configuration, which at 
large transverse momenta behaves as
\begin{equation}
\psi_N (z_1, z_2, z_3; \, \bm{p}_{\perp 1}, \bm{p}_{\perp 2}, 
\bm{p}_{\perp 3}) 
\;\; \propto \;\; 
\sum_{i\neq j} \frac{\Phi_N (z_1, z_2, z_3)}
{p_{\perp i}^2 \; p_{\perp j}^2}.
\end{equation}
Here $\Phi_N$ is the nucleon distribution amplitude, whose
asymptotic shape is $\Phi_N \propto z_1 z_2 z_3$. 
The differential cross section is given by
\begin{eqnarray}
\frac{d\sigma}{dz_{1} \, dz_{2} \, dz_{3} \, 
d^{2}p_{\perp 1} \, d^{2}p_{\perp 2} \, d^{2}p_{\perp 3}} &=&
 c_N \, \left[ \alpha_{s} x G(x, Q^{2}) \right]^{2}
\frac{\Phi^2_N (z_1, z_2, z_3)}{p_{\perp 1}^{4} \,
p_{\perp 2}^{4} \, p_{\perp 3}^{4}} \; F^2_{g}(x, t) \nonumber \\
&\times& \delta^{(2)} 
(\sum \bm{p}_{\perp i} - \bm{\Delta}_\perp ) \; \delta(\sum z_{i}-1) ,
\end{eqnarray}
where $\bm{\Delta}_\perp$ is the momentum transfer to the 
surviving proton ($\Delta_\perp^2 = -t$). The coefficient $c_N$ is 
calculable in QCD. $F_g (x, t)$ denotes the two--gluon form factor of 
the nucleon, see Section~\ref{sec_transverse}. A numerical estimate 
of the cross section for 3--jet production at the LHC energy, 
with $p_\perp$ of one of the jets larger than a given value 
(here: $10\, {\rm GeV}/c$) and all other variables integrated over, gives
\begin{equation}
\sigma(pp\rightarrow {\rm 3jets} + p) \;\; \approx \;\;
\frac{\left[ \alpha_s x G(x,Q^2) \right]^{2}}
{p_\perp^{8}} \;\; \approx \;\; 10^{- (6 \div 7)} 
\left(\frac{10\, {\rm GeV}}{p_\perp}\right)^{8}
\; {\rm mb}.
\end{equation}
The probability of the $|qqq \rangle$ configuration was estimated
using a phenomenological fit to the probability of configurations of
different interaction strengths in a nucleon, see
Refs.~\cite{Frankfurt:1998eu,Frankfurt:1994hf} for details.
An important question is whether the cross section will grow up 
to the LHC energy, as was assumed in this estimate, in which
the gluon density was extrapolated to values of $x \sim 10^{-5}$.

Experimentally, the main difficulty will be to measure jets at 
very high rapidities, $y_{\rm jet}(p_\perp = 10\, {\rm GeV}/c) \sim 6$, 
and with a large background from leading--twist hard diffraction. 
The latter will be suppressed in $pA$ collisions, since the coherent 
3--jet process has much stronger $A$--dependence than the background.

Finally, we note that it would be possible to study also the process 
$pp \rightarrow pp + \mbox{two jets}$, which is similar to pion 
dissociation into two jets. Experimentally, this would require the 
measurement of jets at rapidities $y\sim 4$.
\section{Exclusive diffractive Higgs production}
\label{sec_higgs}
\begin{figure}[b]
\includegraphics[width=10cm,height=7cm]{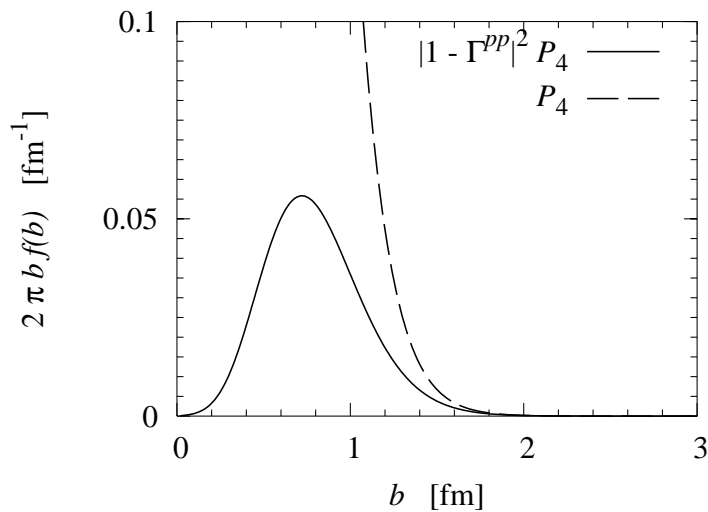}
\caption[]{The impact parameter distribution of the
cross section for diffractive Higgs production at LHC
($s = 14\, {\rm TeV}$). Dashed line: $b$--distribution 
of the hard process, $P_4 (b)$,
Eq.~(\ref{P_4}), {\it cf.}\ Fig.~\ref{fig_pb}. 
Solid line: $b$--distribution of the 
total process, $|1 - \Gamma^{pp} (s, b)|^2 P_4 (b)$.
The mass parameter in the two--gluon form factor
was chosen as $m_g^2 = 1 \, {\rm GeV}^2$.}
\label{fig_survb}
\end{figure}
Hard diffractive processes are also being considered in connection
with the production of new heavy particles in $pp$ collisions at LHC.
In particular, the exclusive diffractive production of Higgs bosons,
\begin{equation}
p + p \;\; \rightarrow \;\; p + \mbox{(gap)} + H + \mbox{(gap)} + p ,
\end{equation}
is regarded as a promising candidate for the Higgs 
search; see Ref.~\cite{Kaidalov:2003fw} and references therein. 
From the point of view of strong interactions, this process involves 
a delicate interplay between ``hard'' and ``soft'' interactions, 
which can be described within our two--scale picture of the transverse
structure of the nucleon \cite{Frankfurt:2004kn}. 
The Higgs boson is produced in a hard partonic 
process, involving the exchange of two hard gluons between the nucleons. 
The impact parameter distribution of the cross section for this process 
is described by the square of the convolution of the transverse spatial
distributions of gluons in the in and out states, $P_4 (b)$, 
defined in Eq.~(\ref{P_4}), where the scale is of the order of the 
gluon transverse momentum squared, 
$\sim M_H^2 / 4$. In addition, the soft interactions between the 
the spectator systems have to conspire in such a way as not to 
fill the rapidity gaps left open by the hard process. 
The probability for this to happen is approximately given by one minus 
the probability of an inelastic $pp$ interaction at a given 
impact parameter, Eq.~(\ref{winel}), or $|1 - \Gamma^{pp} (s, b)|^2$.
The product of the two probabilities, which determines the 
$b$--distribution for the total process, is shown in Fig.~\ref{fig_survb}.
At small $b$ the probability for no inelastic interaction is very small
$|1 - \Gamma^{pp}|^2 \approx 0$, leading to a strong suppression of small
$b$ in the overall distribution. 

\begin{figure}[t]
\includegraphics[width=10cm,height=7cm]{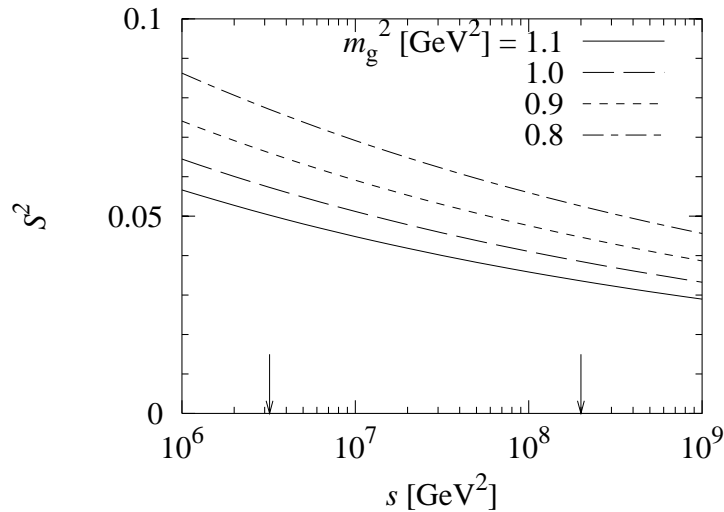}
\caption[]{The rapidity gap survival probability, $S^2$, 
Eq.~(\ref{surv}), as estimated in Ref.~\cite{Frankfurt:2004kn}.
Shown is the result as a function of $s$, for various values 
of the mass parameter in the two--gluon form factor, $m_g^2$. 
The Tevatron and LHC energies are marked by arrows.}
\label{fig_surv}
\end{figure}
The so-called rapidity gap survival probability, which measures the 
``price'' to be paid for leaving the protons intact, is given by 
the integral \cite{Frankfurt:2004kn}
\begin{equation}
S^2 \;\; \equiv \;\; \int d^2 b \; |1 - \Gamma^{pp} (s, b)|^2 \; P_4 (b) ,
\label{surv}
\end{equation}
Fig.~\ref{fig_surv} shows our result for this quantity, with 
$s$ ranging between Tevatron and LHC energies, for various values of the
dipole mass in the two--gluon form factor of the nucleon, $m_g^2$, 
Eq.~(\ref{dipole}). The survival probability decreases with $s$ because the 
size of the ``black'' region at small impact parameters (in which 
inelastic interactions happen with high probability) grows with the 
collision energy. Note that the effective $x$ values 
in the gluon distribution decrease with the energy
(for fixed mass of the produced Higgs boson), resulting in smaller 
effective values of $m_g^2$. This makes the actual drop 
of the survival probability with energy slower than appears from
the fixed--$m_g^2$ curves of Fig.~\ref{fig_surv}.

Our estimates of $S^2$ are in reasonable agreement with those obtained by 
Khoze et al.\ \cite{Khoze:2000wk} in a multi--Pomeron model,
as well as with those reported by Maor et al.\ \cite{Maor}. 
In view of the different theoretical input to these approaches 
this is very encouraging. Note that our estimate of the survival
probability applies equally well to the production of two hard dijets 
instead of a Higgs boson.

\section{Inclusive hard diffractive processes}
Inclusive hard diffractive processes, such as 
\begin{equation}
\begin{array}{lcll}
p + p &\rightarrow& p + \mbox{(gap)} + \mbox{2 jets} + X &
\mbox{``single diffractive''}, 
\\[1ex]
p + p &\rightarrow& p + \mbox{(gap)} + \mbox{2 jets} + X + \mbox{(gap)} + p 
& \mbox{``double diffractive''},
\end{array}
\end{equation}
offer a possibility to probe the ``periphery'' of the proton
with hard scattering processes. The cross section
for these processes is again suppressed compared to the naive 
estimate based on the diffractive parton densities of the proton
measured in $ep$ scattering at HERA. 
As in the case of exclusive diffractive Higgs production, the
cause of this is the very small probability for the nucleons 
not to interact inelastically at small impact parameters.
The suppression factors can be estimated by generalizing
the approach to the description of hard and soft interactions outlined 
in Section~\ref{sec_higgs}. Simple estimates along the lines of 
Eq.~(\ref{surv}) naturally reproduce the suppression factors of the 
order $0.1 \div 0.2$ observed at Tevatron. However, the results in this 
case are more sensitive to the details of the impact parameter dependence 
of the hard scattering process and the soft spectator interactions.
 
In particular, the study of inclusive hard diffractive processes 
at LHC will allow to {\it i)} investigate how the overall 
increase of the nucleon size with energy leads to a suppression of 
hard diffraction, {\it ii)} check how the rate of suppression 
depends on the $x$--value of the parton involved in the hard process, 
{\it iii)} look for the breakdown of Regge factorization, that is, 
the change of the diffractive parton distributions with 
$x_{\pomeron}$.
\section{Conclusions}
Interactions close to the black body limit
play an important role in $pp$ collisions at LHC energies.
They largely determine the strength of the elastic $pp$ amplitude 
at small impact parameters, and lead to a strong suppression of 
hard and soft diffractive processes. They also determine the 
dynamics of spectator interactions in events with new particle 
production, and cause substantial changes in the final state 
characteristics as compared to generic events. A systematic study of 
these effects ({\it e.g.}\ with a trigger on hard dijet production)
will allow one to investigate the small--$x$ dynamics of high gluon 
densities in $pp$ scattering, at densities comparable to those reached
in the heavy ion collisions at LHC energies.

The final state characteristics of central collisions at LHC 
are predicted to be very different from the naive extrapolation of 
Tevatron results, where the effect of ``black interactions'' is
much weaker.

Dedicated studies of hard diffraction at LHC, with a  
large acceptance in the very forward region, can provide
unique information about the transverse spatial distribution of gluons
in the nucleon, as well as the growth of the gluon density with energy.
This information will be crucial for understanding QCD dynamics in 
the regime of high gluon densities, and for observing new phenomena 
reflecting the three--dimensional structure of the nucleon.
\\[1ex]
{\small This work is supported by U.S.\ Department of Energy Contract
DE-AC05-84ER40150, under which the Southeastern Universities Research 
Association (SURA) operates the Thomas Jefferson 
National Accelerator Facility.
L.~F.\ and M.~S.\ acknowledge support by the Binational Scientific 
Foundation. The research of M.~S. was supported by DOE.}

\end{document}